\documentclass[longauth, traditabstract]{aa} 

\usepackage{graphicx}
\usepackage{txfonts}
\usepackage{natbib}
\begin{document}

\title{\emph{Herschel} observations of EXtra-Ordinary Sources: The Terahertz spectrum of Orion KL seen at high spectral resolution\thanks{\emph{Herschel} is an ESA space observatory with science instruments provided by European-led Principal Investigator consortia and with important participation from NASA.}}

\author{
N.~R.~Crockett\inst{1},
E.~A.~Bergin\inst{1},
S.~Wang\inst{1},
D.~C.~Lis\inst{2},
T.~A.~Bell\inst{2},
G.~A.~Blake\inst{2},
A.~Boogert\inst{16},
B.~Bumble\inst{13},
S.~Cabrit\inst{24},
E.~Caux\inst{4,5},
C.~Ceccarelli\inst{6},
J.~Cernicharo\inst{7},
C.~Comito\inst{8},
F.~Daniel\inst{7,9},
M.-L.~Dubernet\inst{10,11},
M.~Emprechtinger\inst{2},
P.~Encrenaz\inst{9},
E.~Falgarone\inst{25}, 
M.~Gerin\inst{9},
T.~F.~Giesen\inst{12},
J.~R.~Goicoechea\inst{7},
P.~F.~Goldsmith\inst{13},
H.~Gupta\inst{13},
R.~G\"usten\inst{8},
P.~Hartogh\inst{22},
F.~Helmich\inst{20},
E.~Herbst\inst{14},
N.~Honingh\inst{12},
C.~Joblin\inst{4,5},
D.~Johnstone\inst{15},
A.~Karpov\inst{2},
J.~H.~Kawamura\inst{13},
J.~Kooi\inst{2},
J.-M.~Krieg\inst{9},
W.~D.~Langer\inst{13},
W.D.~Latter\inst{16},
S.~D.~Lord\inst{16},
S.~Maret\inst{6},
P.~G.~Martin\inst{17},
G.~J.~Melnick\inst{18},
K.~M.~Menten\inst{8},
P.~Morris\inst{16},
H.~S.~P.~M\"uller\inst{12},
J.~A.~Murphy\inst{19},
D.~A.~Neufeld\inst{3},
V.~Ossenkopf\inst{12,20},
J.~C.~Pearson\inst{13},
M.~P\'erault\inst{9},
T.~G.~Phillips\inst{2},
R.~Plume\inst{21},
S.-L.~Qin\inst{12},
P.~Roelfsema\inst{20},
R.~Schieder\inst{12},
P.~Schilke\inst{8,12},
S.~Schlemmer\inst{12},
J.~Stutzki\inst{12},
F.~F.~S.~van der Tak\inst{21},
A.~Tielens\inst{23},
N.~Trappe\inst{19},
C.~Vastel\inst{4,5},
H.~W.~Yorke\inst{13},
S.~Yu\inst{13},
\and
J.~Zmuidzinas\inst{2}
}
\institute{Department of Astronomy, University of Michigan, 500 Church Street, Ann Arbor, MI 48109, USA\\ 
\email{ncrocket@umich.edu}
\and California Institute of Technology, Cahill Center for Astronomy and Astrophysics 301-17, Pasadena, CA 91125 USA
\and  Department of Physics and Astronomy, Johns Hopkins University, 3400 North Charles Street, Baltimore, MD 21218, USA
\and Centre d'\'etude Spatiale des Rayonnements, Universit\'e de Toulouse [UPS], 31062 Toulouse Cedex 9, France
\and CNRS/INSU, UMR 5187, 9 avenue du Colonel Roche, 31028 Toulouse Cedex 4, France
\and Laboratoire d'Astrophysique de l'Observatoire de Grenoble, 
BP 53, 38041 Grenoble, Cedex 9, France.
\and Centro de Astrobiolog\'ia (CSIC/INTA), Laboratiorio de Astrof\'isica Molecular, Ctra. de Torrej\'on a Ajalvir, km 4
28850, Torrej\'on de Ardoz, Madrid, Spain
\and Max-Planck-Institut f\"ur Radioastronomie, Auf dem H\"ugel 69, 53121 Bonn, Germany 
\and LERMA, CNRS UMR8112, Observatoire de Paris and \'Ecole Normale Sup\'erieure, 24 Rue Lhomond, 75231 Paris Cedex 05, France
\and LPMAA, UMR7092, Universit\'e Pierre et Marie Curie,  Paris, France
\and  LUTH, UMR8102, Observatoire de Paris, Meudon, France
\and I. Physikalisches Institut, Universit\"at zu K\"oln,
              Z\"ulpicher Str. 77, 50937 K\"oln, Germany
\and Jet Propulsion Laboratory,  Caltech, Pasadena, CA 91109, USA
\and Departments of Physics, Astronomy and Chemistry, Ohio State University, Columbus, OH 43210, USA
\and National Research Council Canada, Herzberg Institute of Astrophysics, 5071 West Saanich Road, Victoria, BC V9E 2E7, Canada 
\and Infrared Processing and Analysis Center, California Institute of Technology, MS 100-22, Pasadena, CA 91125
\and Canadian Institute for Theoretical Astrophysics, University of Toronto, 60 St George St, Toronto, ON M5S 3H8, Canada
\and Harvard-Smithsonian Center for Astrophysics, 60 Garden Street, Cambridge MA 02138, USA
\and  National University of Ireland Maynooth. Ireland
\and SRON Netherlands Institute for Space Research, PO Box 800, 9700 AV, Groningen, The Netherlands
\and Department of Physics and Astronomy, University of Calgary, 2500
University Drive NW, Calgary, AB T2N 1N4, Canada
\and MPI f\"ur Sonnensystemforschung, D 37191 Katlenburg-Lindau,
Germany
\and Leiden Observatory, Leiden University, PO Box 9513, NL-2300 RA Leiden, The Netherlands
\and LERMA \& UMR8112 du CNRS, Observatoire de Paris, 61, Av. de l'Observatoire, 75014 Paris, France
\and LERMA, CNRS UMR8112, Observatoire de Paris and \'Ecole Normale Sup\'erieure, 24 Rue Lhomond, 75231 Paris Cedex 05, France
}


\abstract{We present the first high spectral resolution observations of Orion KL in the frequency ranges 1573.4 -- 1702.8~GHz (bands 6b) and 1788.4 -- 1906.8~GHz (band 7b) obtained using the HIFI instrument on board the \emph{Herschel} Space Observatory. We characterize the main emission lines found in the spectrum, which primarily arise from a range of components associated with Orion KL including the hot core, but also see widespread emission from components associated with molecular outflows traced by H$_{2}$O, SO$_{2}$, and OH.  We find that the density of observed emission lines is significantly diminished in these bands compared to lower frequency \emph{Herschel}/HIFI bands.}
   \titlerunning{THz Spectrum of Orion KL}
	\authorrunning{Crockett et al.}
   \maketitle
%

\section{Introduction}

The Kleinmann-Low nebula within the Orion molecular cloud (Orion KL) is the best studied massive star forming region in the Milky Way. This region is characterized by a high IR luminosity \citep{kleinmann67} and rich molecular line emission.  As such, it has been the subject of numerous molecular line surveys in the millimeter and submillimeter that have characterized its mm/sub-mm wave spectrum \citep[see e.g.][and references therein]{schilke97, comito05, tercero10}. These surveys reveal the presence of a prodigious variety of molecular species in addition to several distinct spatial/velocity components \citep[i.e. the hot core, compact ridge, plateau, and extended ridge;][]{blake87,persson07}. These observations provide insight into the complex chemical and physical processes that characterize how massive stars form and interact with their natal environment. Because Orion KL is the closest such massive star forming region ($\sim$414 pc; \citealt{menten07}), it is an ideal choice for further inquiry in understanding the chemistry and physics of the gas in close proximity to these stars.

Although the subject of much spectroscopic study in the mm /sub-mm ($\lambda \gtrsim$ 300$\mu$m) during the past 30 years, high resolution observations at Terahertz (THz) frequencies of Orion KL have been unavailable from the ground due to atmospheric absorption. ISO provided the first comprehensive spectroscopic view of Orion KL at these wavelengths. \citet{lerate06} presented spectroscopic observations in the wavelength range 44 -- 188 $\mu$m (1.6 -- 6.8~THz) with a resolving power of $\lambda/\Delta \lambda$ $\sim$ 6800-9700 using the long wavelength spectrometer (LWS) on board ISO. These data showed a spectrum dominated by emission from H$_{2}$O, OH, and CO, but little or no emission from more complex species (such as methanol, methyl formate, dimethyl ether, etc.), which litter the spectrum at submillimeter wavelengths.

The HIFI instrument \citep{degraauw10} on board the \emph{Herschel} Space Observatory \citep{pilbratt10} provides the first opportunity to characterize the THz spectrum with high spectral resolution and sensitivity. In this Letter, we present the first high resolution ($\lambda/\Delta \lambda$ $\sim$ $10^{6}$) spectrum of Orion KL above 1.57 THz obtained using the HIFI instrument. These observations, taken as part of the guaranteed time key program \emph{Herschel observations of EXtra-Ordinary Sources: The Orion and Sagittarius B2 Starforming Regions} (HEXOS), are able to probe the chemical inventory and  kinematic structure of Orion KL at an unprecedented level. In this work , we characterize the high resolution THz spectrum. We further demonstrate and discuss why the observed line density is reduced when compared to lower frequencies.

\begin{figure*}
\label{spec67}
\centering
\includegraphics[width=1.62\columnwidth]{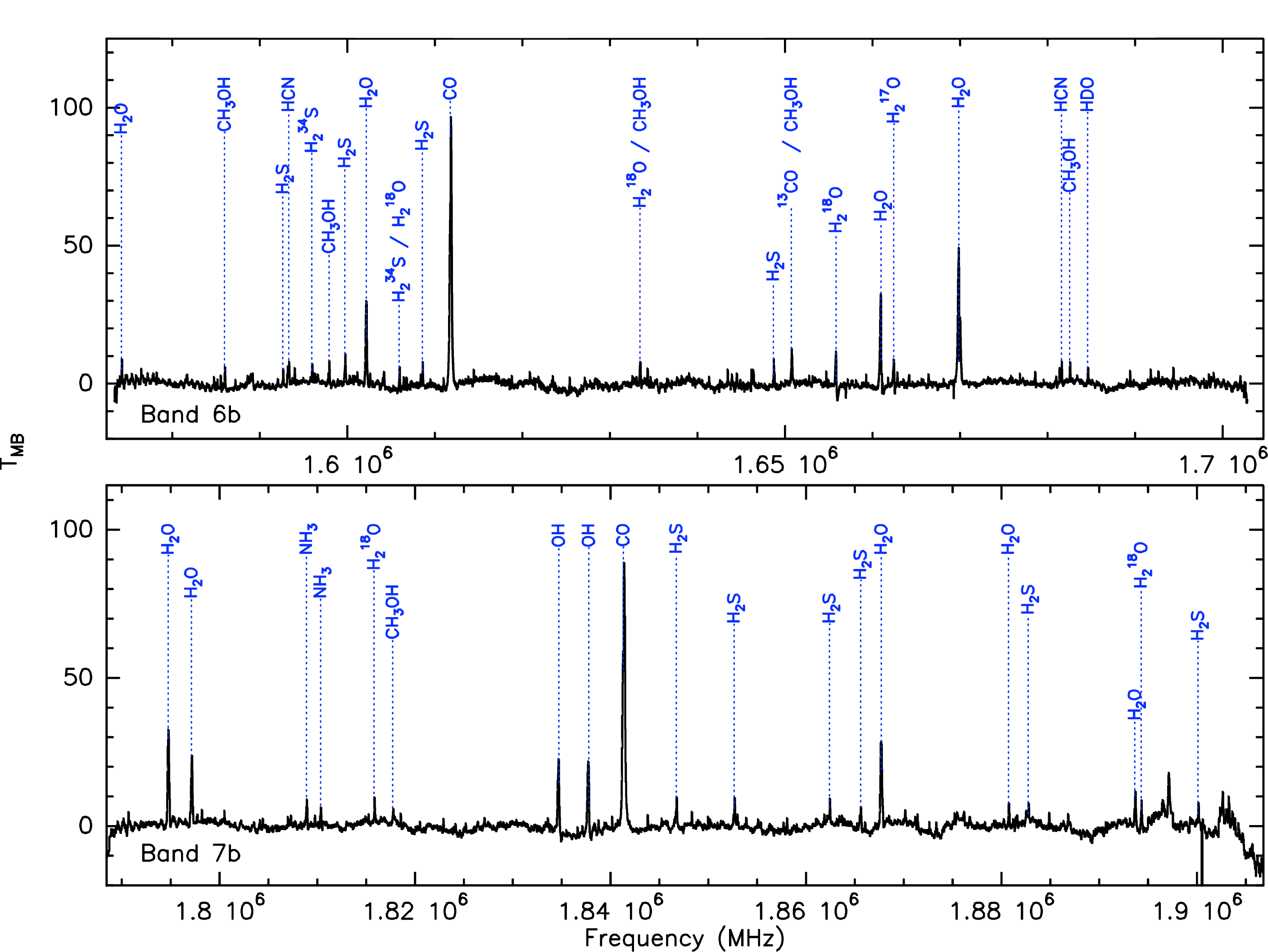}
\caption{SSB spectrum of the Orion KL hot core in Bands 6b (top panel) and 7b (bottom panel) smoothed to a velocity resolution of $\sim$4.5~km s$^{-1}$. The strongest lines (T$_{MB}$ $\gtrsim$ 7 K) are labelled.}
\end{figure*}

\section{Observations}
\label{obs}

The observations were carried out on March 22-23, 2010 using the wide band spectrometer (WBS) with a spectral resolution of 1.1 MHz (0.19~km s$^{-1}$ at 1.75 THz) over a $\sim$2.4~GHz IF bandwidth. The data were taken in dual beam switch (DBS) mode using the fast chop setting pointed towards the Orion hot core at coordinates $\alpha_{J2000} = 5^h35^m14.5^s$ and $\delta_{J2000} = -5^{\circ}22'30.9''$.  The beam size at 1.75~THz is 12$^{\prime\prime}$ and the DBS reference beams lie approximately 3$^{\prime}$ east and west. Both H and V polarization data were obtained. However, we only present the H polarization here because the mixer is optimized for the H polarization. These data were reduced using HIPE \citep{ott10} with pipeline version 2.4.

The observations presented here are full spectral scans of Bands 6b and 7b, meaning they cover a frequency range of 1573.4 - 1702.8 GHz (176.2 - 190.7 $\mu$m) and 1788.4 - 1906.8 GHz (157.3 - 167.7 $\mu$m), respectively. These spectral scans consist of double-sideband (DSB) spectra with a redundancy of 4, which are deconvolved into single-sideband (SSB) spectra. This procedure is outlined in \citet{bergin10}. We applied the standard HIFI deconvolution using the {\it doDeconvolution} task within HIPE with no channel weighting or gain correction. Strong spurs and noisy DSB data sets were not included in the deconvolution and no fringing correction was applied.  All data presented in this Letter are deconvolved SSB spectra.

After the deconvolution was performed, the data were exported to FITS format and all subsequent data reduction and analysis was performed using the IRAM GILDAS package. Main beam efficiencies for bands 6b and 7b were assumed to be 0.64 and 0.63, respectively. We estimate the typical RMS in both bands to be $T_A^* \approx$ 0.9~K at the original spectral resolution.  

\section{Results}

\begin{figure}[h]
\centering
\includegraphics[width=1.0\columnwidth]{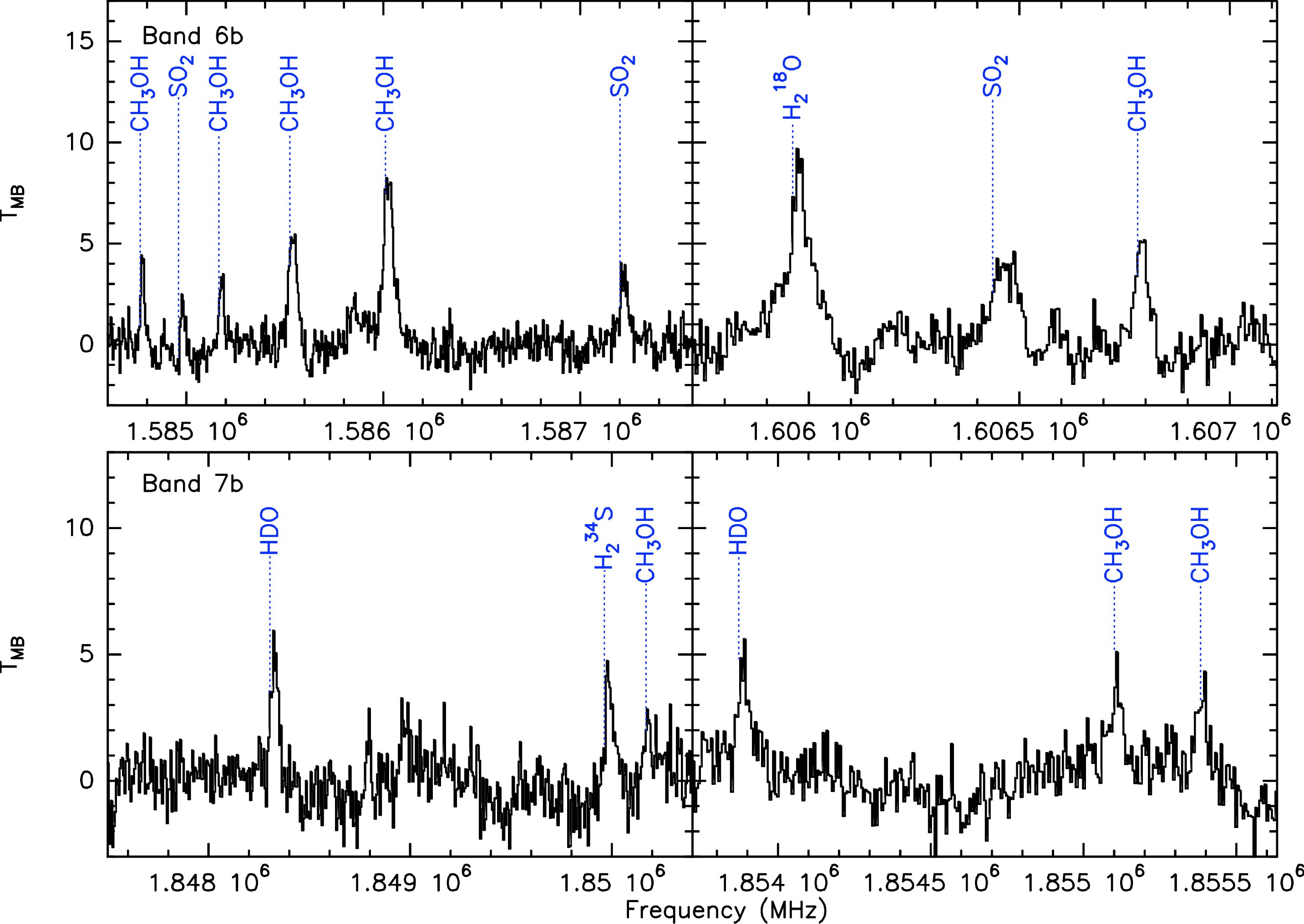}
\caption{A small sample of weak lines in Bands 6b (top panels) and 7b (bottom panels). All spectra are smoothed to a resolution of $\sim$1~km s$^{-1}$.}
\label{weak}
\end{figure}

The SSB spectra for Bands 6b and 7b are given in Fig.~\ref{spec67} smoothed to a velocity resolution of $\sim$4.5 km s$^{-1}$ and corrected for a V$_{LSR}$ = 9~km s$^{-1}$ with the most prominent lines (peak T$_{MB}$ $\gtrsim$ 7~K) labelled. Polynomial baselines of order 2 are also subtracted from each spectrum. We find that these observations are dominated by strong lines of CO, H$_{2}$O, and OH as was reported by \citet{lerate06}. With the higher spectral resolution of HIFI, we also  detect additional strong lines of CH$_{3}$OH, H$_{2}$S, HCN, and HDO.  Line identifications were made with the aid of the XCLASS program\footnote{http://www.astro.uni-koeln.de/projects/schilke/XCLASS}  which accesses both the CDMS \citep[][http://www.cdms.de]{muller01, muller05} and JPL \citep[][ http://spec.jpl.nasa.gov]{pickett98} molecular databases. We list these transitions along with their integrated intensities in Table~\ref{table1}.  Line intensities were measured using the CLASS data reduction and analysis software package. In instances where there were blends, Gaussian profiles were fit to the lines and the results from the fitted profiles are reported; otherwise the total intensity is measured directly using the BASE command. All line intensities were measured using spectra smoothed to a velocity resolution of $\sim$1 km s$^{-1}$. Uncertainties in the integrated intensities, $\sigma_{I}$, were computed using the relation $\sigma_{I}$(K km s$^{-1}$) =$\sqrt{N}(\delta v) RMS$ where $\delta v$ is the resolution in velocity space, N is the number of channels over which the intensity is measured, and RMS is the root mean square deviation in the vicinity of the line. In addition to the lines listed in Table~\ref{table1}, we also detect many additional weak transitions of CH$_{3}$OH,  SO$_{2}$, H$_{2}$S, and H$_{2}$O along with their isotopologues. Examples of several weak lines detected in Bands 6b and 7b are plotted in Fig.~\ref{weak}. Integrated line intensities for these weaker transitions along with peak intensities for all lines will be reported in a later study.

When comparing these spectra to other lower frequency HIFI bands, it is readily apparent that the line density is significantly diminished when compared to the lower frequency bands \citep[see e.g.][]{bergin10,wang10}. We estimate that the total fraction of channels taken up by lines is $\sim$0.23 in the lower frequency bands compared to $\sim$0.07 in Bands 6 and 7. We reached these estimates by counting the number of channels in emission in the frequency ranges 858.1 -- 958.1 GHz (Band 3b) and 1788.4 -- 1898.5 GHz (Band 7b). We adopt these line density estimates as being representative of the low and high frequency bands, respectively. Although not formally presented in this Letter, a full spectral scan of Orion KL taken in Band 3b was also obtained as part of the HEXOS key program and used to estimate the line density here. These data were reduced in the same way as Bands 6b/7b. Both bands were smoothed to a velocity resolution of $\sim$1 km s$^{-1}$ and any channel that had a value T$_{MB}$ $>$ 2.5~K (after baseline subtraction) was flagged as being in emission in 7b. This threshold is approximately what we have estimated as 3$\times$ the RMS in T$_{MB}$ in Band 7b at a resolution of 1~km s$^{-1}$ (RMS~$\sim$~0.8~K). Because the beam size, $\theta$, decreases as a function of frequency ($\theta_{3b}$$\sim$24$^{\prime\prime}$), this value was scaled to an equivalent RMS in Band 3b using the following relation,
\begin{equation}
T_{RMS_{3b}}=T_{RMS_{7b}}\left( \frac{\theta_{7b}} {\theta_{3b}} \right)^{2},
\end{equation}
which assumes that the source is significantly smaller than both beam sizes. Thus the reduced beam size should be more coupled to the smaller spatial components (e.g. the hot core). One might therefore naively expect the line density to increase at THz frequencies. The opposite trend, however, is observed.

\begin{table}
\caption{Strong Lines in Band 6b and 7b}
\label{table1}
\begin{tabular}{l c r r c}
\hline \hline
Molecule	&	Frequency	&	Transition	&	$\int{T_{MB}dv}$	&   Notes\\
		&	(MHz)		&			&	(K km s$^{-1}$)		 &   \\
\hline
		&				&	Band 6b	&					&		\\
\hline
H$_{2}$O			&	1574232.073	&	6$_{4,3}$ -- 7$_{1,6}$	& 70.8      $\pm$ 6.7		&	 \\
CH$_{3}$OH		&	1586012.991	&	8$_{5,1}$ -- 7$_{4,1}$	& 85.3      $\pm$ 3.0		& 3	\\
				&	1586013.008	&	8$_{5,0}$ -- 7$_{4,0}$	&					& 3	\\
H$_{2}$S			&	1592669.425	&	7$_{2,5}$ -- 7$_{1,6}$	& 69.5      $\pm$ 3.4		&	 \\
HCN				&	1593341.504	&	18--17				& 138.4    $\pm$ 5.5		& 1	 \\
H$_{2}$$^{34}$S	&	1595984.323	&	4$_{2,3}$ -- 3$_{1,2}$	& 76.3      $\pm$ 5.1		& 	 \\
CH$_{3}$OH		&	1597947.024	&	9$_{6,0}$ -- 8$_{5,0}$	& 84.4      $\pm$ 4.1		& 3	 \\
				&	1597947.024	&	9$_{6,1}$ -- 8$_{5,1}$	&					& 3	\\
H$_{2}$S			&	1599752.748	&	4$_{2,3}$ -- 3$_{1,2}$	& 258.0    $\pm$ 6.4		&	 \\
H$_{2}$O			&	1602219.182	&	4$_{1,3}$ -- 4$_{0,4}$	& 959.0    $\pm$ 8.5		&	 \\
H$_{2}$$^{34}$S	&	1605957.883	&	6$_{1,5}$ -- 6$_{0,6}$	& 116.0    $\pm$ 6.0		& 3	 \\
H$_{2}$$^{18}$O	&	1605962.460	&	4$_{1,3}$ -- 4$_{0,4}$	&					& 3	 \\ 
H$_{2}$S			&	1608602.794	&	6$_{2,5}$ -- 6$_{1,6}$	& 91.8      $\pm$ 4.2		&	 \\
CO				&	1611793.518	&	14 -- 13				& 4653.0 $\pm$ 12.0	& 2	 \\ 
H$_{2}$$^{18}$O	&	1633483.600	&	2$_{2,1}$ -- 2$_{1,2}$	& 192.0   $\pm$ 9.3		& 3	 \\
CH$_{3}$OH		&	1633493.496	&	13$_{4,0}$ -- 12$_{3,0}$	& 					& 3	\\
H$_{2}$S			&	1648712.816	&	4$_{2,2}$ -- 3$_{3,1}$	& 124.9    $\pm$ 4.9		& 1	 \\
$^{13}$CO		&	1650767.302	&	15 -- 14				& 287.0    $\pm$ 8.0		& 3	 \\
CH$_{3}$OH		&	1650817.827	&	21$_{4,0}$ -- 20$_{3,0}$	& 					& 3	 \\
H$_{2}$$^{18}$O	&	1655867.627	&	2$_{1,2}$ -- 1$_{0,1}$	& 75.6      $\pm$ 8.2		& 2	 \\
H$_{2}$O			&	1661007.637	&	2$_{2,1}$ -- 2$_{1,2}$	& 1008.0 $\pm$ 9.2		& 2	 \\
H$_{2}$$^{17}$O	&	1662464.387	&	2$_{1,2}$ -- 1$_{0,1}$	& 155.0   $\pm$ 7.4		& 2	 \\
H$_{2}$O			&	1669904.775	&	2$_{1,2}$ -- 1$_{0,1}$	& 2266.0 $\pm$ 10.4	& 2	 \\
HCN				&	1681615.473	&	19 -- 18				& 136.0   $\pm$ 5.8		&	 \\
CH$_{3}$OH		&	1682556.723	&	10$_{5,1}$ -- 9$_{4,1}$	& 76.6     $\pm$ 3.9		& 3	 \\
				&	1682556.856	&	10$_{5,0}$ -- 9$_{4,0}$	&					& 3	\\
HDO				&	1684605.824	&	6$_{1,5}$ -- 6$_{0,6}$	& 71.9     $\pm$ 4.2		&	 \\
\hline
		&				&	Band 7b	&					&		\\
\hline
H$_{2}$O			&	1794788.953	&	6$_{2,4}$ -- 6$_{1,5}$	&	969.0    $\pm$ 8.5	&	 \\
H$_{2}$O			&	1797158.762	&	7$_{3,4}$ -- 7$_{2,5}$	&	648.0    $\pm$ 4.9	&	 \\
NH$_{3}$			&	1808935.550	&	3$_{1,1}$ -- 2$_{1,0}$	&	158.0    $\pm$ 6.4	& 2	 \\
NH$_{3}$			&	1810377.792	&	3$_{2,1}$ -- 2$_{2,0}$	&	50.1      $\pm$ 8.3	& 2	 \\
H$_{2}$$^{18}$O	&	1815853.411	&	5$_{3,2}$ -- 5$_{2,3}$	&	91.0      $\pm$ 5.5	&	 \\
CH$_{3}$OH		&	1817752.285	&	7$_{7,0}$ -- 6$_{6,0}$	&	35.7      $\pm$ 3.2	&	 \\
OH				&	1834747.350	&	$^{2}\Pi_{1/2}3/2^{-}$ -- 1/2$^{+}$	&	627.0   $\pm$ 9.7	& 2, 4 \\ 
OH				&	1837816.820	&	$^{2}\Pi_{1/2}3/2^{+}$ -- 1/2$^{-}$	&	640.0   $\pm$ 11.5	& 2, 4 \\   
CO				&	1841345.506	&	16 -- 15				   &	3820.0 $\pm$ 10.4	& 2	 \\
H$_{2}$S			&	1846768.559	&	6$_{1,6}$ -- 5$_{0,5}$	   &	237.0   $\pm$ 5.8	& 	 \\ 
H$_{2}$S			&	1852685.693	&	5$_{1,4}$ -- 4$_{2,3}$	&	181.0   $\pm$ 4.8	&	 \\
H$_{2}$S			&	1862435.697	&	5$_{2,4}$ -- 4$_{1,3}$	&	84.7     $\pm$ 5.2	&	 \\
H$_{2}$S			&	1865620.670	&	3$_{3,0}$ -- 2$_{2,1}$	&	173.6   $\pm$ 7.8	& 	 \\
H$_{2}$O			&	1867748.594	&	5$_{3,2}$ -- 5$_{2,3}$	&	864.0   $\pm$ 8.8	&	 \\
H$_{2}$O			&	1880752.750	&	6$_{3,4}$ -- 7$_{0,7}$	&	135.0   $\pm$ 5.4	&	 \\
H$_{2}$S			&	1882773.396	&	8$_{3,6}$ -- 8$_{2,7}$	&	37.9     $\pm$ 4.0	&	 \\
H$_{2}$O			&	1893686.801	&	3$_{3,1}$ -- 4$_{0,4}$	&	265.0  $\pm$ 5.9	&	 \\
H$_{2}$$^{18}$O	&	1894323.823	&	3$_{2,2}$ -- 3$_{1,3}$	&	161.0  $\pm$ 6.4	&	 \\
H$_{2}$S			&	1900140.572	&	7$_{1,6}$ -- 7$_{0,7}$	&	83.80  $\pm$ 6.7	& 3	 \\
				&	1900177.906	&	7$_{2,6}$ -- 7$_{1,7}$	&					& 3	 \\ 
\hline
\end{tabular}
\caption{Notes: 1~--~Line intensities were fit using Gaussian profiles because of a blend. 2~--~Lines contained self absorption. 3~--~Subsequent lines with this note were severely blended and could not be separated by line fitting. The same integrated intensity is reported for both entries. 4~--~Other OH transitions contributed to the integrated intensity reported. The strongest transition is given.}
\end{table}

\section{Discussion}
\label{sec:disc}

\begin{figure}
\centering
\includegraphics[width=1.0\columnwidth]{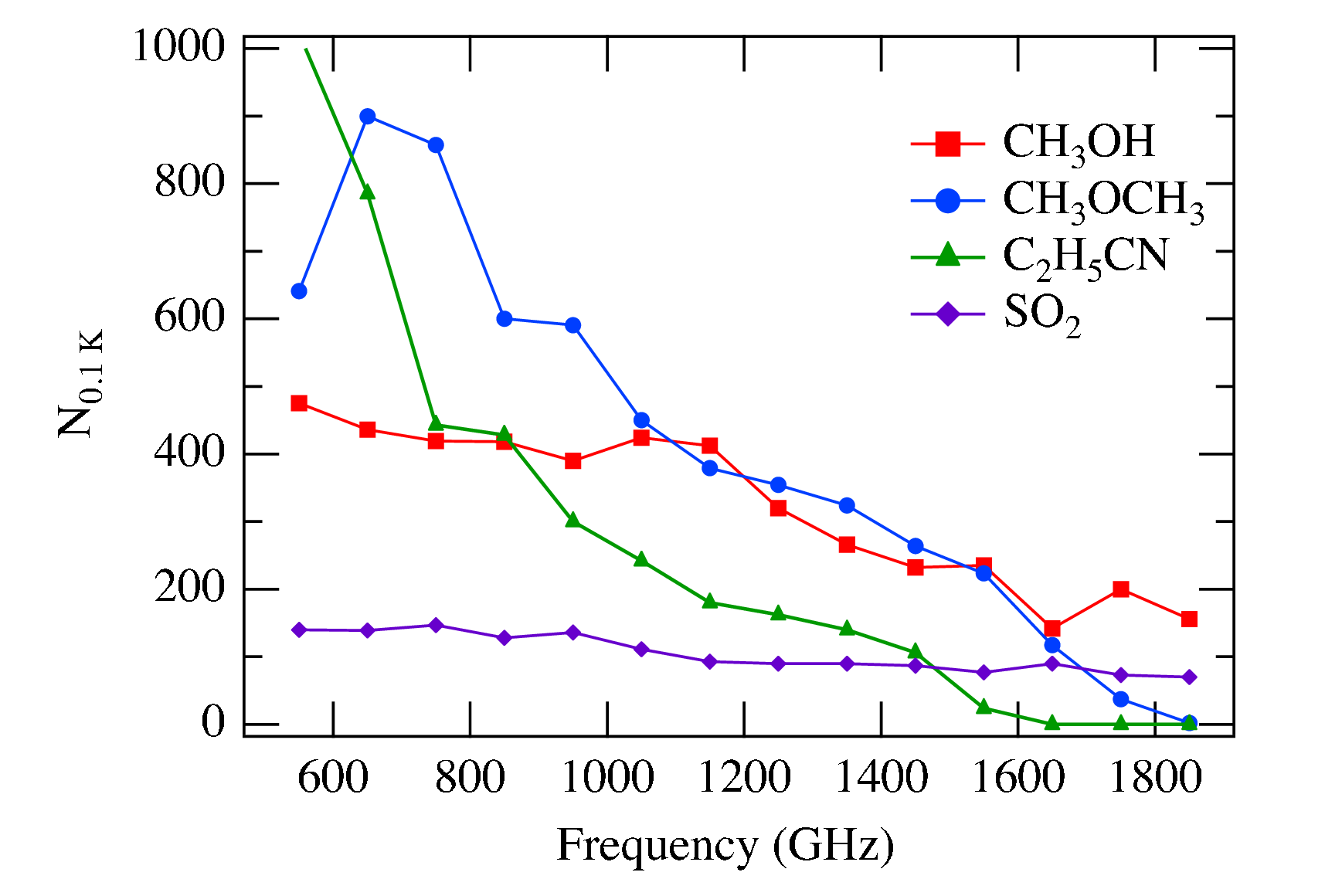}
\caption{Predicted number of lines with peak emission $>$ 0.1~K based on an LTE model in 100 GHz bins for select "weeds" as a function of frequency.}
\label{linedensity}
\end{figure}

One of the primary reasons for the reduced line density in the high frequency bands is the fall off in emission from complex organic molecules -- in particular the ``weeds'' such as CH$_3$OCH$_3$, SO$_2$, C$_2$H$_5$CN, and, of course, CH$_3$OH. In Fig.~\ref{linedensity} we present the number of emissive lines for select ``weeds'' as a function of frequency.   To estimate these numbers we assumed LTE and predicted the emission for each species assuming T $= 150$K.  We use the total column estimated for each molecule on the basis of \citet{comito05} and in addition assumed a velocity width of 5 km s$^{-1}$.   If the predicted emission was above 0.1 K then we counted the line as potentially emissive in our 100 GHz bins.  In this fashion we counted $N_{0.1K}$, which is shown in the figure. As can be seen, there is a general decrease in emission for all species but its particularly evident for CH$_3$OCH$_3$ and C$_2$H$_5$CN.  CH$_3$OH has a small factor of 2 decrease in the number of lines and, at the zeroth level, this is seen in our data which has numerous weak methanol transitions scattered throughout the band. 

Another possibility is that the dust emission from the hot core is optically thick in the high frequency bands; thus the dust would absorb all of the photons emitted from the molecules in the hot core. To explore this more closely we can examine the dust opacity expected within the hot core itself. 
Plume et al. (2010, in preparation) used multiple transitions of C$^{18}$O and spectrally isolated the hot core. They estimate an N(C$^{18}$O) = 1.7$\times$10$^{16}$ cm$^{-2}$ which yields a total H$_{2}$ column of 3.4$\times$10$^{23}$ cm$^{-2}$ assuming n(C$^{18}$O)/n(H$_{2}$) = 1.7$\times$10$^{-7}$ \citep{frerking82}. Using the relation given in \citet[][Equation 10]{hildebrand83}, we estimate a $\tau$$\sim$0.1 at 171 $\mu$m, putting it slightly lower than being optically thick.  

It is clear that there are other emission components in this region as we see widespread emission from a variety of molecules in the high frequency bands. 
However, we still observe many molecules (CH$_{3}$OH, H$_{2}$O, HDO, and HCN) that have velocity components in their spectral profiles that are coincident to those expected from the hot core and other components (e.g. the outflows). If these emission components do arise in the the hot core, it is likely that the molecular emission region must lie in front of any optically thick core. Given the presence of strong physical gradients in the density and temperature profiles \citep{wright96,blake96} and the fact that the dust is marginally optically thick, this is not unrealistic.

A final contributor to the decrease in the line emission could be non-LTE excitation. At high frequencies there are a larger number of high excitation lines which could be more difficult to excite even at densities of 10$^{6}$ -- 10$^{7}$ cm$^{-3}$. This needs to be more directly calculated using a molecule such as CH$_3$OH with collision rates that extend to temperatures greater than 200 -- 300~K.

\section{Summary}

We have characterized the high frequency spectrum of Orion KL. We find that the spectrum is dominated by strong lines of CO, H$_{2}$O, HDO, OH, CH$_{3}$OH, H$_{2}$S, HCN, and NH$_{3}$. We also detect many weaker transitions of CH$_{3}$OH, H$_{2}$O, HDO, and SO$_{2}$. We find that the line density is diminished in the high frequency bands when compared to the lower frequency bands and provide a number of explanations as to why this may be. 

\begin{acknowledgements}
HIFI has been designed and built by a consortium of institutes and university departments from across 
Europe, Canada and the United States under the leadership of SRON Netherlands Institute for Space
Research, Groningen, The Netherlands and with major contributions from Germany, France and the US. 
Consortium members are: Canada: CSA, U.Waterloo; France: CESR, LAB, LERMA,  IRAM; Germany: 
KOSMA, MPIfR, MPS; Ireland, NUI Maynooth; Italy: ASI, IFSI-INAF, Osservatorio Astrofisico di Arcetri- 
INAF; Netherlands: SRON, TUD; Poland: CAMK, CBK; Spain: Observatorio Astron—mico Nacional (IGN), 
Centro de Astrobiolog'a (CSIC-INTA). Sweden:  Chalmers University of Technology - MC2, RSS \& GARD; 
Onsala Space Observatory; Swedish National Space Board, Stockholm University - Stockholm Observatory; 
Switzerland: ETH Zurich, FHNW; USA: Caltech, JPL, NHSC.
Support for this work was provided by NASA through an award issued by JPL/Caltech.
CSO is supported by the NSF, award AST-0540882.
\end{acknowledgements}

\bibliographystyle{aa}

\begin{thebibliography}{20}
\expandafter\ifx\csname natexlab\endcsname\relax\def\natexlab#1{#1}\fi

\bibitem[{{Bergin} {et~al.}(2010, this volume){Bergin}}]{bergin10}
{Bergin}, E.~A., {Phillips}, T.~G., {Comito}, C., {et~al.} 2010, \aap, this volume

\bibitem[{{Blake} {et~al.}(1996){Blake}, {Mundy}, {Carlstrom}, {Padin},
  {Scott}, {Scoville}, \& {Woody}}]{blake96}
{Blake}, G.~A., {Mundy}, L.~G., {Carlstrom}, J.~E., {et~al.} 1996, \apjl, 472,
  L49+

\bibitem[{{Blake} {et~al.}(1987){Blake}, {Sutton}, {Masson}, \&
  {Phillips}}]{blake87}
{Blake}, G.~A., {Sutton}, E.~C., {Masson}, C.~R., \& {Phillips}, T.~G. 1987,
  \apj, 315, 621

\bibitem[{{Comito} {et~al.}(2005){Comito}, {Schilke}, {Phillips}, {Lis},
  {Motte}, \& {Mehringer}}]{comito05}
{Comito}, C., {Schilke}, P., {Phillips}, T.~G., {et~al.} 2005, \apjs, 156, 127

\bibitem[{{de Graauw} {et~al.}(2010){de Graauw}, {Helmich}, {Phillips},
  {Stutzki}, {Caux}, {Whyborn}, {Dieleman}, {Roelfsema}, {Aarts}, {Assendorp},
  {Bachiller}, {Baechtold}, {Barcia}, {Beintema}, {Belitsky}, {Benz}, {Bieber},
  {Boogert}, {Borys}, {Bumble}, {Ca{\"i}s}, {Caris}, {Cerulli-Irelli},
  {Chattopadhyay}, {Cherednichenko}, {Ciechanowicz}, {Coeur-Joly}, {Comito},
  {Cros}, {de Jonge}, {de Lange}, {Delforges}, {Delorme}, {den Boggende},
  {Desbat}, {Diez-Gonz{\'a}lez}, {di Giorgio}, {Dubbeldam}, {Edwards},
  {Eggens}, {Erickson}, {Evers}, {Fich}, {Finn}, {Franke}, {Gaier}, {Gal},
  {Gao}, {Gallego}, {Gauffre}, {Gill}, {Glenz}, {Golstein}, {Goulooze},
  {Gunsing}, {G{\"u}sten}, {Hartogh}, {Hatch}, {Higgins}, {Honingh}, {Huisman},
  {Jackson}, {Jacobs}, {Jacobs}, {Jarchow}, {Javadi}, {Jellema}, {Justen},
  {Karpov}, {Kasemann}, {Kawamura}, {Keizer}, {Kester}, {Klapwijk}, {Klein},
  {Kollberg}, {Kooi}, {Kooiman}, {Kopf}, {Krause}, {Krieg}, {Kramer},
  {Kruizenga}, {Kuhn}, {Laauwen}, {Lai}, {Larsson}, {Leduc}, {Leinz}, {Lin},
  {Liseau}, {Liu}, {Loose}, {L{\'o}pez-Fernandez}, {Lord}, {Luinge}, {Marston},
  {Mart{\'{\i}}n-Pintado}, {Maestrini}, {Maiwald}, {McCoey}, {Mehdi}, {Megej},
  {Melchior}, {Meinsma}, {Merkel}, {Michalska}, {Monstein}, {Moratschke},
  {Morris}, {Muller}, {Murphy}, {Naber}, {Natale}, {Nowosielski}, {Nuzzolo},
  {Olberg}, {Olbrich}, {Orfei}, {Orleanski}, {Ossenkopf}, {Peacock}, {Pearson},
  {Peron}, {Phillip-May}, {Piazzo}, {Planesas}, {Rataj}, {Ravera}, {Risacher},
  {Salez}, {Samoska}, {Saraceno}, {Schieder}, {Schlecht}, {Schl{\"o}der},
  {Schm{\"u}lling}, {Schultz}, {Schuster}, {Siebertz}, {Smit}, {Szczerba},
  {Shipman}, {Steinmetz}, {Stern}, {Stokroos}, {Teipen}, {Teyssier}, {Tils},
  {Trappe}, {van Baaren}, {van Leeuwen}, {van de Stadt}, {Visser}, {Wildeman},
  {Wafelbakker}, {Ward}, {Wesselius}, {Wild}, {Wulff}, {Wunsch}, {Tielens},
  {Zaal}, {Zirath}, {Zmuidzinas}, \& {Zwart}}]{degraauw10}
{de Graauw}, T., {Helmich}, F.~P., {Phillips}, T.~G., {et~al.} 2010, \aap, 518,
  L6+

\bibitem[{{Frerking} {et~al.}(1982){Frerking}, {Langer}, \&
  {Wilson}}]{frerking82}
{Frerking}, M.~A., {Langer}, W.~D., \& {Wilson}, R.~W. 1982, \apj, 262, 590

\bibitem[{{Hildebrand}(1983)}]{hildebrand83}
{Hildebrand}, R.~H. 1983, \qjras, 24, 267

\bibitem[{{Kleinmann} \& {Low}(1967)}]{kleinmann67}
{Kleinmann}, D.~E. \& {Low}, F.~J. 1967, \apjl, 149, L1+

\bibitem[{{Lerate} {et~al.}(2006){Lerate}, {Barlow}, {Swinyard}, {Goicoechea},
  {Cernicharo}, {Grundy}, {Lim}, {Polehampton}, {Baluteau}, {Viti}, \&
  {Yates}}]{lerate06}
{Lerate}, M.~R., {Barlow}, M.~J., {Swinyard}, B.~M., {et~al.} 2006, \mnras,
  370, 597

\bibitem[{{Menten} {et~al.}(2007){Menten}, {Reid}, {Forbrich}, \&
  {Brunthaler}}]{menten07}
{Menten}, K.~M., {Reid}, M.~J., {Forbrich}, J., \& {Brunthaler}, A. 2007, \aap,
  474, 515

\bibitem[{{M{\"u}ller} {et~al.}(2005){M{\"u}ller}, {Schl{\"o}der}, {Stutzki},
  \& {Winnewisser}}]{muller05}
{M{\"u}ller}, H.~S.~P., {Schl{\"o}der}, F., {Stutzki}, J., \& {Winnewisser}, G.
  2005, Journal of Molecular Structure, 742, 215

\bibitem[{{M{\"u}ller} {et~al.}(2001){M{\"u}ller}, {Thorwirth}, {Roth}, \&
  {Winnewisser}}]{muller01}
{M{\"u}ller}, H.~S.~P., {Thorwirth}, S., {Roth}, D.~A., \& {Winnewisser}, G.
  2001, \aap, 370, L49

\bibitem[{{Ott}(2010){Ott}}]{ott10}
{Ott}, S. 2010, in ASP Conference Series, Astronomical Data Analysis Software and Systems XIX,Y. Mizumoto, K.~I. Morita, and M.Ohishi, eds., in press 

\bibitem[{{Persson} {et~al.}(2007){Persson}, {Olofsson}, {Koning}, {Bergman},
  {Bernath}, {Black}, {Frisk}, {Geppert}, {Hasegawa}, {Hjalmarson}, {Kwok},
  {Larsson}, {Lecacheux}, {Nummelin}, {Olberg}, {Sandqvist}, \&
  {Wirstr{\"o}m}}]{persson07}
{Persson}, C.~M., {Olofsson}, A.~O.~H., {Koning}, N., {et~al.} 2007, \aap, 476,
  807

\bibitem[{{Pickett} {et~al.}(1998){Pickett}, {Poynter}, {Cohen}, {Delitsky},
  {Pearson}, \& {Muller}}]{pickett98}
{Pickett}, H.~M., {Poynter}, I.~R.~L., {Cohen}, E.~A., {et~al.} 1998, Journal
  of Quantitative Spectroscopy and Radiative Transfer, 60, 883

\bibitem[{{Pilbratt} {et~al.}(2010){Pilbratt}, {Riedinger}, {Passvogel},
  {Crone}, {Doyle}, {Gageur}, {Heras}, {Jewell}, {Metcalfe}, {Ott}, \&
  {Schmidt}}]{pilbratt10}
{Pilbratt}, G.~L., {Riedinger}, J.~R., {Passvogel}, T., {et~al.} 2010, \aap,
  518, L1+

\bibitem[{{Schilke} {et~al.}(1997){Schilke}, {Groesbeck}, {Blake}, \&
  {Phillips}}]{schilke97}
{Schilke}, P., {Groesbeck}, T.~D., {Blake}, G.~A., \& {Phillips}, T.~G. 1997,
  \apjs, 108, 301

\bibitem[{{Tercero} {et~al.}(2010){Tercero}, {Cernicharo}, {Pardo}, \&
  {Goicoechea}}]{tercero10}
{Tercero}, B., {Cernicharo}, J., {Pardo}, J.~R., \& {Goicoechea}, J.~R. 2010,
  ArXiv e-prints

\bibitem[{{Wang} {et~al.}(2010, in press){Wang}}]{wang10}
{Wang}, S., {Bergin}, E.~A., {Crockett}, N.~R., {et~al.} 2010, \aap, in press

\bibitem[{{Wright} {et~al.}(1996){Wright}, {Plambeck}, \& {Wilner}}]{wright96}
{Wright}, M.~C.~H., {Plambeck}, R.~L., \& {Wilner}, D.~J. 1996, \apj, 469, 216

\end{thebibliography}

\end{document}